\documentclass[12pt]{iopart}
\usepackage{epsfig}

\newcommand{\mean}[1]{\langle #1 \rangle}
\newcommand{\rms}[1]{\langle #1^2 \rangle^{1/2}}

\newcommand{\centps}[2]{
    \begin{center}
        \epsfig{file=#1,width=#2mm}
    \end{center}
}

\begin{document}
\jl{1}

\letter{Alternating steady state in one-dimensional flocking}

\author{O J O'Loan and M R Evans}

\address{Department of Physics and Astronomy, University of Edinburgh,
        Mayfield Road, Edinburgh EH9 3JZ, Scotland}

\begin{abstract}
We study flocking in one dimension, introducing a lattice model in which
particles can move either left or right. We find that the model exhibits a
continuous nonequilibrium phase transition from a condensed phase, in which a
single `flock' contains a finite fraction of the particles, to a homogeneous
phase; we study the transition using numerical finite-size scaling.
Surprisingly, in the condensed phase the steady state is alternating, with the
mean direction of motion of particles reversing stochastically on a timescale
proportional to the logarithm of the system size. We present a simple argument
to explain this logarithmic dependence. We argue that the reversals are
essential to the survival of the condensate. Thus, the discrete directional
symmetry is not spontaneously broken.
\end{abstract}


\vspace{1cm}

Flocking---the collective motion of a large number of self-propelled
entities---is a behaviour exhibited by many living beings such as birds, fish
and bacteria. Generically, flocks are driven, non-equilibrium systems with many
degrees of freedom and as such they have recently attracted much attention in
the physics community \cite{VCBCS,CSV,TT-long,BDG,CBV,Albano,SSMHS}.

Vicsek et al.~\cite{VCBCS} introduced a simple microscopic model in dimensions
$d \ge 2$ in which particles move with a constant speed; they interact only by
tending to align with their neighbours. Simulations \cite{CSV} and a continuum
theory \cite{TT-long} showed the existence of a low-noise ordered phase in
which the mean velocity of the particles is non-zero, i.e., a phase in which
the rotational symmetry of the model is spontaneously broken.

Flocking in one dimension (1d) is less relevant to biological systems than in
higher dimensions, but is nevertheless interesting from a fundamental
viewpoint. The models studied to date in $d \ge 2$ possess a continuous
(rotational) symmetry. The 1d case is necessarily different---since particles
are constrained to move either left or right on a line, the underlying
directional symmetry is discrete. Czir\'ok et al.~\cite{CBV} have introduced an
off-lattice model of 1d flocking in which particles with continuous velocity
variables move on a line. Particles tend to move in the same direction as their
neighbours but `errors' are made due to the presence of noise.  Simulations and
a continuum theory indicate that a continuous phase transition occurs as the
noise or particle density is varied \cite{CBV}, the ordered phase being
characterised by the presence of a large `flock'.

When a phase transition occurs from a disordered to an ordered phase, it is
usually accompanied by spontaneous symmetry breaking; in the ordered phase, the
spontaneously broken symmetry is accompanied by ergodicity breaking in the
thermodynamic limit. In an ordered system with discrete symmetry, such as the
Ising model in $d \ge 2$, one expects the `flip time'---the time the system
takes to move from one symmetry-related state to another---to diverge
exponentially with system size.

In this work we study a simple lattice model of flocking in 1d. We show, using
finite-size scaling, that the model has a continuous phase transition from a
homogeneous to a condensed phase. However, the condensed phase is not
symmetry-broken, but {\em alternating}---we find that the mean particle
velocity alternates its sign on a timescale which grows only logarithmically
with system size; moreover, these reversals are essential to the maintenance of
the order.

We now define the model that we study. We consider $N$ particles on a periodic
1d lattice of $L$ sites. The particle density is $\rho = N/L$. Each particle,
labelled by $\mu$, has a position $x_{\mu} \in \{1,2,\ldots,L\}$ and a velocity
(or direction) $v_{\mu} = \pm 1$. To update the system, a particle is chosen at
random. Its velocity $v_{\mu}$ is flipped with probability $W_{\mu}$ and then
$x_{\mu} \to x_{\mu} + v_{\mu}$. The flip probability $W_{\mu}$ is given by
\begin{equation}
	W_{\mu} =
		\left[ 1 - (1-2\eta) v_{\mu} U(x_{\mu}) \right]/2,
\end{equation} 
where $U(y)$ is the velocity of the majority of the particles (including
particle $\mu$) at site $y$ and its two nearest neighbours; we take $U = 0$
when there is no majority. In brief, particles acquire the velocity
of the majority of their neighbours with probability $1-\eta$.

While our lattice model is somewhat simpler than the off-lattice model of
reference \cite{CBV}, it is conceptually similar. However, whereas the
off-lattice model has only one source of noise (a random perturbation added to
the velocity of a particle at each time step), the noise in our model has two
distinct sources. The first is due to the flipping of particle velocities and
its strength is parameterised by $\eta$; the second is due to the random
sequential dynamics of the model. Thus, unlike the off-lattice model, our
model does not become deterministic in the limit $\eta \to 0$. 

\begin{figure}
	\centps{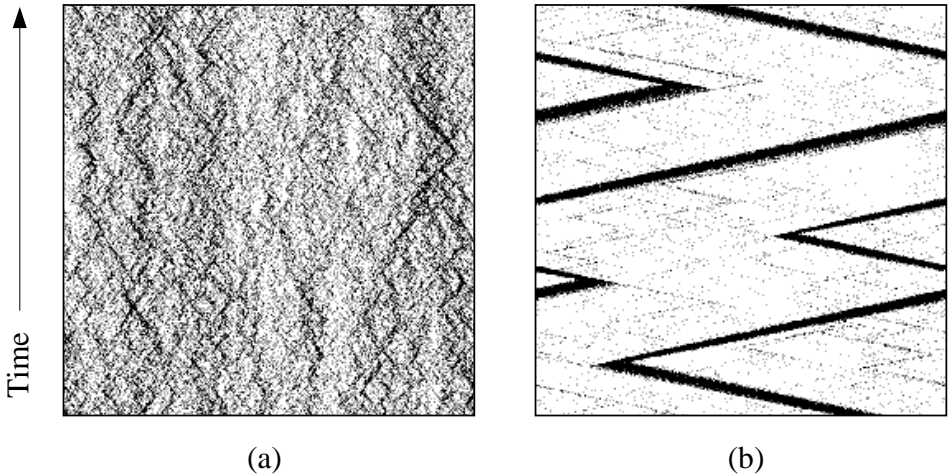}{125}
	\caption{Space-time plots of steady-state systems with (a) $\eta=0.2$ and
	(b) $\eta=0.02$. In both cases, $\rho=1$ and $L=1000$.  A darker grey level
	indicates a higher particle density. There is one timestep between each
	snapshot in (a) and 5 in (b).}
	\label{fig:ST}
\end{figure}

Before showing that the model exhibits a continuous phase transition, we first
discuss the  qualitative behaviour. When $\eta$ is large, the
system has a fairly homogeneous steady state, illustrated in figure
\ref{fig:ST}(a) by a space-time plot for $\eta = 0.2$. However, when $\eta$ is
small, `condensation' occurs---a large fraction of the particles are contained
in a single large flock, which we refer to as the {\it condensate}. A
space-time plot of the steady state for $\eta = 0.02$ is shown in figure
\ref{fig:ST}(b). The direction of motion of the condensate reverses at fairly
regular intervals.  These reversals are sometimes, but not always, triggered by
a collision with a smaller flock travelling in the opposite direction. Just
after the condensate has reversed its direction, it is very dense. It then
gradually spreads out, becoming more diffuse at the front than at the back,
before reversing again. The steady state clearly has no time-reversal symmetry
and is {\it alternating}, by which we mean that the system reverses its mean
velocity at stochastic intervals.  A quantity measuring the spatial order in
the system, such as the variance of the number of particles per site, decreases
gradually as the condensate spreads, then increases suddenly as the condensate
`flips', before decaying again.

If a condensate were never to reverse its direction of motion, it would
eventually become so diffuse that it would cease to exist---the alternating
character of the steady state is essential for the existence of the
condensate. The condensate has no preferred direction of motion and we conclude
that directional symmetry is not broken.

\begin{figure}
	\centps{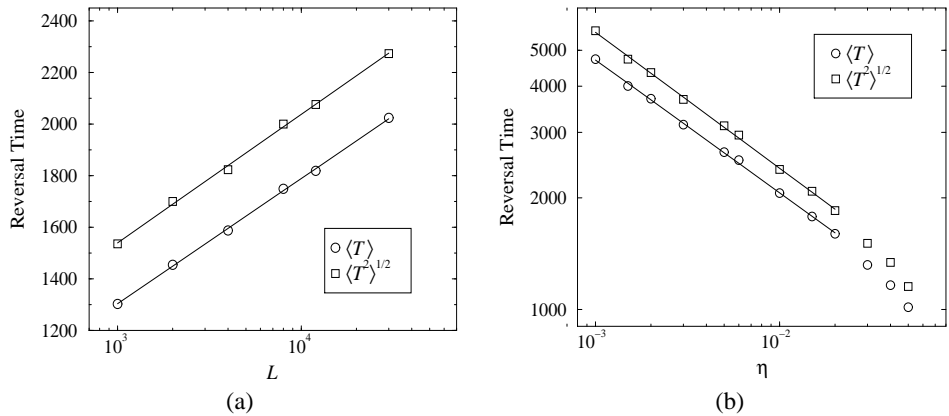}{125} 
	\caption{
	(a) Log-linear plot of $\mean{T}$ and $\rms{T}$ as a function
	of $L$ for $\eta=0.02$ and $\rho=1$. The straight lines are fits to the
	data.
	(b) Log-log plot of $\mean{T}$ and $\rms{T}$ as a function of
	$\eta$ for $L=4000$ and $\rho=1$. The solid lines are fits to the data for
	$\eta \le 0.02$ with exponents of $-0.358 \pm 0.01$ and $-0.366 \pm 0.01$
	respectively.}
	\label{fig:Flipping1}
\end{figure}

We now discuss the behaviour of the condensate in more detail. Let $p(T)$ be
the probability that the condensate reverses its direction (i.e.~that the mean
particle velocity changes sign) a time $T$ after the previous reversal.
Figure~\ref{fig:Flipping1}(a) shows a log-linear plot of $\mean{T}$ and
$\rms{T}$ as a function of system size $L$ for fixed $\eta = 0.02$. Both are
proportional to $\log L$. Thus, while no spontaneous symmetry breaking occurs,
the mean time between reversals does diverge weakly in the thermodynamic limit,
as does the variance.

\begin{figure}
	\centps{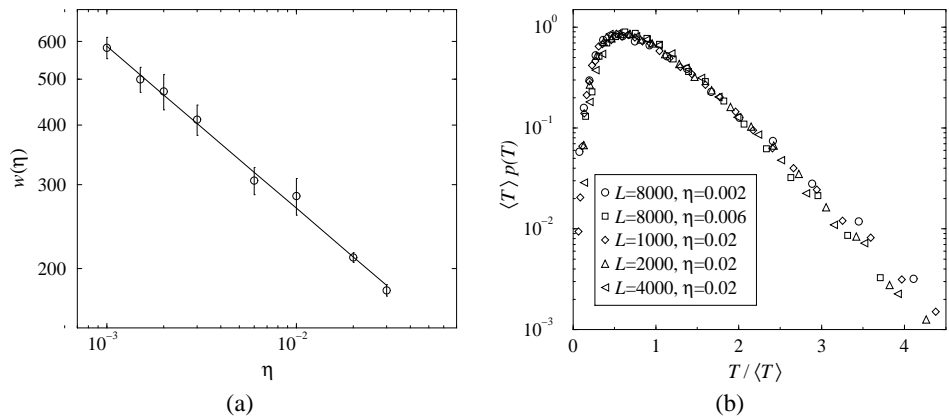}{125} 
	\caption{
	(a) Log-log plot of $w(\eta)$ for $\rho=1$. The solid line is a fit to the
	data.
	(b) Linear-log plot of the scaled reversal time distribution $\mean{T}
	p(T)$ as a function of scaled time $T/\mean{T}$ for $\rho=1$ and various
	values of $L$ and $\eta$.}
	\label{fig:Flipping2}
\end{figure}

Figure~\ref{fig:Flipping1}(b) shows a log-log plot of $\mean{T}$ and $\rms{T}$
as a function of $\eta$ for fixed $L=4000$. For $\eta$ less than about $0.02$,
both $\mean{T}$ and $\rms{T}$ diverge as power laws in $\eta$. The data
suggests that the exponent, which we call $\lambda$, is the same in both cases
with a value of $-0.36 \pm 0.01$. We have seen in figure~\ref{fig:Flipping1}(a)
that $\mean{T} \propto \log L$ for fixed $\eta$ and $L$ large; we denote the
constant of proportionality by $w(\eta)$. Figure~\ref{fig:Flipping2}(a) shows a
log-log plot of $w(\eta)$ and a least-squares fit to the data gives an estimate
of $-0.34 \pm 0.02$ for the slope; it seems likely that this is equal to
$\lambda$ and so we conclude that $\mean{T} \sim \eta^{\lambda} \log L$ for
large $L$ and small $\eta$.

The fact that $\langle T \rangle$ and $\langle T^2 \rangle^{1/2}$ are both
proportional to $\log L$ for fixed $\eta$, and both proportional to
$\eta^{\lambda}$ for fixed $L$ suggests that the distribution of reversal times
may have the scaling form $p(T) \sim f(T/\mean{T}) / \mean{T}$ asymptotically
for large $L$ and small $\eta$. Figure~\ref{fig:Flipping2}(b) shows the scaled
distribution for various values of $L$ and $\eta$---the data collapse is
fairly good. 

We now present a simple argument which may explain why a condensate containing
$N$ particles flips direction in a time of order $\log N$ (or equivalently
$\log L$ if the density is fixed) for large $N$. Consider a system in which,
at $t=0$, all $N$ particles are positioned at $x=0$ and have velocity $+1$. Let
$q(x,t)$ be the probability that a given particle in this initial flock has
position $x$ at time $t$. In the limit $\eta \to 0$, each particle hops forward
with probability $1/N$ in a time $\Delta t$ so that in the limit $N \to
\infty$, $\Delta t = 1/N$, the distribution $q(x,t)$ is Poisson:
\begin{equation}
	q(x,t) = \frac{e^{-t}\;t^x}{x!}.  \label{eqn:Poisson}
\end{equation}
For $\eta$ nonzero, this distribution will be modified by the flipping of
particle velocities. For example, particles will leave the back of the
condensate when $\eta$ is nonzero but here we neglect this effect.

We now ask the following question: if some perturbation or fluctuation (due to
collision with a small flock or the spontaneous flipping of particles at the
front of the condensate) should cause all the particles in the condensate
having position $x > z$ to flip velocity, will this in turn cause the particles
with position $x = z$ to flip? We make the approximation that the particles
with $x > z$ form a `shock' moving back through the condensate so that at time
$t$ they all have position $z+1$ and velocity $-1$. Then the particles with
position $z$ will flip if $R(z,t) > 1$, where
\begin{equation}
	R(z,t) = \frac{1}{q(z-1,t) + q(z,t)} \sum_{x=z+1}^{\infty} q(x,t) =
	\frac{t}{z+t} \sum_{x=1}^{\infty} 
				\frac{z! \; t^x}{(z+x)!}.
\end{equation}
For a given position $z$ we define a critical time $t_c(z)$ through the
solution of $R(z,t_c)=1$ so that for $t > t_c(z)$, the particles at position
$z$ are susceptible to flipping.

It is easy to see that $R(z,t+1) > R(z+1,t)$ so that if the particles at $z+1$
are susceptible to flipping (i.e.~if $R(z+1,t) > 1$) then those at $z$ are
subsequently also susceptible. Hence, the reversal time of the entire
condensate is determined by the time one must wait for the leading particles to
flip and it suffices to find $t_c(z)$ in the limit $z \to \infty$. It is
straightforward to show that $R(z,t) \to s^2/(1-s^2)$ as $z \to \infty$ with
$t=sz$; this implies that $t_c(z) \to z/\sqrt{2}$. Therefore, $t_c$ is
approximately proportional to $z$ for large $z$.

For a condensate containing a large but finite number of particles $N$, we
estimate $z^*(N,t)$, the position of the leading edge of the condensate,
through the expression $q(z^*,t) = 1/N$. For the Poisson distribution when $t
\propto z^*$ we find that $z^* \sim \log N$. Therefore our estimate for the 
time at which the condensate becomes susceptible to reversal is $t_c(z^*)$
which, being proportional to $z^*$, is of the order of $\log N$.

Since this simple analysis makes the assumption that one can decouple the
spreading of the condensate and the fluctuations which may cause it to reverse
its direction, it makes no predictions about the non-trivial $\eta$ dependence
of the reversal time and the scaling behaviour of the distribution of reversal
times.

\begin{figure}
	\centps{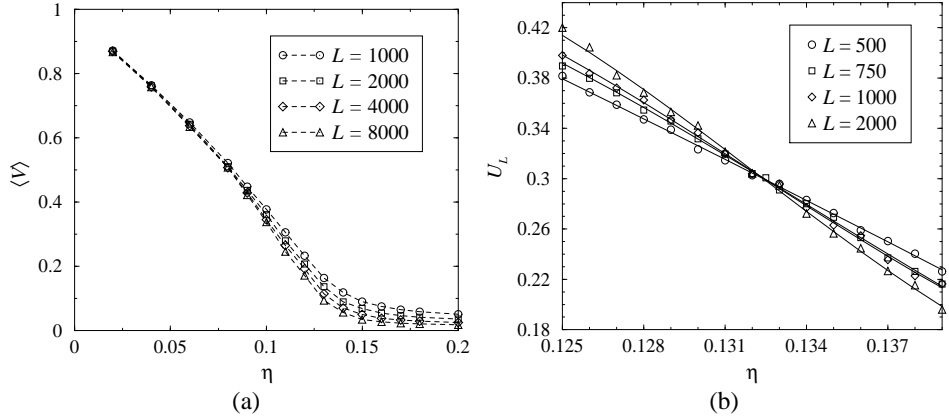}{125} 
	\caption{(a) $\langle V \rangle$ as a function of $\eta$ for various system
	sizes with $\rho = 1$. (b) $U_L(\eta)$ for various system sizes in the
	vicinity of $\eta_c$, given by the common intersection point. The solid
	lines are cubic polynomial fits to the data.}
	\label{fig:FS1}
\end{figure}

This completes our discussion of the condensed phase; we now turn to a
numerical finite-size scaling analysis \cite{BH} of the phase transition
between the condensed phase and the homogeneous phase for fixed $\rho = 1$. We
have performed Monte Carlo simulations for system sizes between $L=50$ and
$L=2000$, averaging over between $5 \times 10^6$ and $2 \times 10^7$ time steps
in the steady state for each set of parameters. We have found that the
absolute value of the mean particle velocity $V$, defined by
\begin{equation}
	V = \frac{1}{N} \left| \sum_{\mu=1}^{N} v_{\mu} \right|,
	\label{eqn:V-def}
\end{equation}
is a convenient order parameter since, between reversals in the condensed
phase, the majority of particles have the same velocity, that of the
condensate. During the long (but $\mathcal{O}(\log L)$) time intervals between
reversals, $V$ fluctuates about some well-defined mean value. Figure
\ref{fig:FS1}(a) shows a plot of $\langle V \rangle$ against $\eta$ for several
different system sizes. The crossover from the disordered to the ordered regime
becomes sharper with increasing system size, suggesting the presence of a
continuous phase transition.

In calculating critical quantities, it is useful to define both the `static
susceptibility' and the fourth-order cumulant \cite{BH}, given respectively by
\begin{equation}
	\chi(L) = L [ \langle V^2 \rangle_L - \langle V \rangle_L^2 ]\;\;;
		\;\;\;\;\;\;\;
	U_L = 1 - \frac{\langle V^4 \rangle_L}{3 \langle V^2 \rangle_L},
	\label{eqn:U-def}
\end{equation}
with the angle brackets indicating steady state time averages. In analogy with
magnetic systems we assume that, near criticality,
$\langle V \rangle \sim (\eta_c - \eta)^\beta$,
$\chi \sim |\eta_c - \eta|^{-\gamma}$ and
$\xi \sim |\eta_c - \eta|^{-\nu}$,
where $\xi$ is the correlation length.

Standard finite-size scaling theory \cite{BH} leads to the following relations
at criticality:
\begin{equation}
	\langle V \rangle_L \sim L^{-\beta/\nu};\;\;\;\;\;\;\;
	\chi(L) \sim L^{\gamma/\nu};\;\;\;\;\;\;\; 
	U_L = U^*.
\end{equation}
The critical point $\eta_c$ can therefore be identified as the value of $\eta$
at which $U_L$ takes on its fixed point value $U^*$. Figure \ref{fig:FS1}(b)
shows that the curves $U_L(\eta)$ do indeed have a common intersection point at
$\eta_c = 0.1325 \pm 0.001$. Also, we find $U^* = 0.30 \pm 0.01$.

\begin{figure}
	\centps{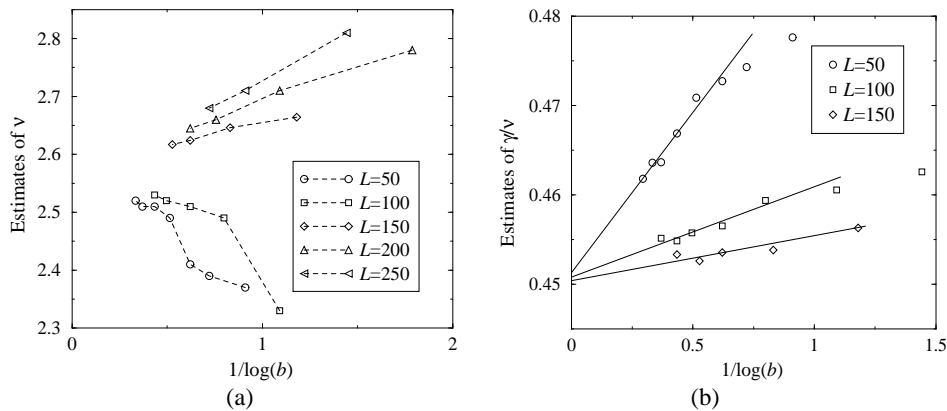}{125} 
	\caption{Estimates of (a) $\nu$ and (b) $\gamma/\nu$, obtained using
	 (\ref{eqn:nucalc}) and (\ref{eqn:gammacalc}),
	plotted against $1/\log b$ for various system sizes. The lines in 
	both are guides to the eye.}
	\label{fig:FS2}
\end{figure}

The slopes of the functions $U_L(\eta)$ at $\eta=\eta_c$ can be used to
calculate the critical exponent $\nu$ since \cite{BH} 
\begin{equation}
	1/\nu = 
	\log \left[ \left.(\partial U_{bL}/{\partial U_L})\right|_{\eta_c}
		\right] / \log b
	\;\; + \;\;\mbox{corrections to scaling}.
	\label{eqn:nucalc}
\end{equation}
We have found that $U_L(\eta)$ is approximately linear
near  $\eta_c$ for $L$ less than about 1000, allowing a
reasonably precise estimate of the derivative in (\ref{eqn:nucalc}). In
figure~\ref{fig:FS2}(a) we plot estimates of $\nu$ obtained using
(\ref{eqn:nucalc}) against $1/\log b$ for various $L$.
Extrapolating $b \to \infty$, thus taking account of corrections to scaling, we
estimate $\nu = 2.57 \pm 0.05$.

The other independent exponent $\gamma/\nu$ is given by
\begin{equation}
	\gamma/\nu = 
	\log [ \chi(bL,\eta_c) / \chi(L,\eta_c)]/ \log b
	\;\; + \;\;\mbox{corrections to scaling}.
	\label{eqn:gammacalc}
\end{equation}
Figure \ref{fig:FS2}(b) shows the resulting estimates of $\gamma/\nu$ plotted
against $1/\log b$ for different values of $L$. As before, we extrapolate $b
\to \infty$ and estimate $\gamma/\nu = 0.451 \pm 0.005$.

The finite-size scaling formalism relies on the fact that the hyperscaling
relation $d = \gamma/\nu + 2\beta/\nu$ holds \cite{BH}.  At $\eta = \eta_c$ one
can calculate $\beta/\nu$ via
\begin{equation}
	\beta/\nu = 
	-\log [ \langle V \rangle_{bL} / \langle V \rangle_L]/ \log b
	\;\; + \;\;\mbox{corrections to scaling}.
	\label{eqn:betacalc}
\end{equation}
We were unable to estimate $\beta/\nu$ as precisely as $\gamma/\nu$ but, within
error bounds, we have found that hyperscaling does indeed hold. Thus, our final
estimates for the critical exponents are $\beta = 0.705 \pm 0.02$, $\gamma =
1.16 \pm 0.04$ and $\nu = 2.57 \pm 0.05$.

In reference~\cite{CBV} $\beta$ was estimated (without finite-size scaling) for
an off-lattice model of 1d flocking to be $0.6 \pm 0.05$. While our value of
$\beta = 0.705 \pm 0.02$ is not in close agreement with this, it is still
possible that both models are in the same universality class. Measurement of
$U^*$ in the off-lattice model would provide a good test of universality.

Above, we have used standard finite-size scaling to study the order parameter
distribution. The success of this approach is surprising when one considers
that the order parameter $V$ does not fully reflect the alternating nature of
the steady state in the condensed phase---the `spreading' of the condensate is
not captured.

We have argued that spontaneous symmetry breaking does not occur in our model
and that the reversal times are $\mathcal{O}(\log L)$.  This contrasts with
equilibrium systems where reversal times exponentially large in $L$ result from
surmounting free energy barriers. It is difficult to see how $\mathcal{O}(\log
L)$ reversal times could be explained in an analogous fashion; the reversal
mechanism is fundamentally nonequilibrium.

Finally, we comment on the effects of imposing a restriction on the local
particle density in our model; this is achieved by setting the maximum number
of particles allowed on a site to be $M$. Not surprisingly, we have found that
for any finite $M<N$, the condensed phase is suppressed; instead, for small
$\eta$, the system forms `domains' of maximally occupied sites. Each of these
domains comprises two competing sub-domains composed of particles with
velocities $+1$ and $-1$. The interface between the two sub-domains performs a
random walk as a result of the flipping of particle velocities and the dynamics
resembles that of a 1d ferromagnet.

\ack
We thank M E Cates for many useful discussions and a critical reading of the
manuscript and N B Wilding for helpful advice on finite-size scaling.


\section*{References}

\begin{thebibliography}{1}

\bibitem{VCBCS}
Vicsek T, Czir\'ok A, Ben-Jacob E, Cohen I and Shochet O 1995 {\em Phys. Rev.
  Lett.\/} {\bf 75} 1226

\bibitem{CSV}
Czir\'ok A, Stanley H~E and Vicsek T 1997 {\em J. Phys. A: Math. Gen.\/} {\bf
  30} 1375

\bibitem{TT-long}
Toner J and Tu Y 1998 {\em Phys. Rev. {\rm E}\/} {\bf 58} 4828

\bibitem{BDG}
Bussemaker H~J, Deutsch A and Geigant E 1997 {\em Phys. Rev. Lett.\/} {\bf 78}
  5018

\bibitem{CBV}
Czir\'ok A, Barab\'asi A and Vicsek T 1997 Preprint cond-mat/9712154

\bibitem{Albano}
Albano E~V 1996 {\em Phys. Rev. Lett.\/} {\bf 77} 2129

\bibitem{SSMHS}
Shimoyama N, Sugawara K, Mizuguchi T, Hayakawa Y and Sano M 1996 {\em Phys.
  Rev. Lett.\/} {\bf 76} 3870

\bibitem{BH}
Binder K and Heermann D~W 1997 {\em {M}onte {C}arlo Simulation in Statistical
  Physics\/}. Solid-State Sciences (Berlin Heidelberg: Springer), 3rd edn.

\end{thebibliography}
\end{document}